\documentclass[floatfix,showpacs,amssymb,amsmath,eqsecnum,superscriptaddress]{aip-cp}

\usepackage[numbers]{natbib}
\usepackage{rotating}
\usepackage{graphicx}
\usepackage{amsfonts}
\usepackage{amsbsy}
\usepackage[usenames]{color}
\usepackage{epstopdf}
\usepackage{subfigure}
\definecolor{red_n}{rgb}{1.0, 0.0, 0.0}
\definecolor{brown_n}{rgb}{0.6, 0.4, 0.2}
\definecolor{cyan_n}{RGB}{0.0, 255.0, 255.0}
\definecolor{blue_n}{rgb}{0.0, 0.0, 1.0}
\definecolor{green_n}{rgb}{0.0, 0.5, 0.0}
\definecolor{orange_n}{RGB}{255.0, 127.0, 0.0}
\definecolor{magenta_n}{RGB}{255.0, 0.0, 255.0}
\definecolor{purple_n}{RGB}{128.0, 0.0, 128.0}
\definecolor{gray_n}{RGB}{128.0, 128.0, 128.0}
\definecolor{dark_green}{rgb}{0.0, 0.5, 0.0}

\begin{document}

\title{Fluctuation-Induced Interactions Between Ellipsoidal Particle and Planar Substrate Immersed in Critical Medium}

\author[aff1]{G. S. Valchev\corref{cor1}}
\author[aff1]{D. M. Dantchev}
\eaddress{daniel@imbm.bas.bg}

\affil[aff1]{Institute of Mechanics, Bulgarian
	Academy of Sciences, Acad. G. Bonchev Str., Block 4, BG-1113 Sofia,	Bulgaria.}
\corresp[cor1]{Corresponding author: gvalchev@imbm.bas.bg}

\maketitle

\begin{abstract}
In our study we investigate the behaviour of the net force (NF) emerging between an ellipsoidal particle and a thick plate (slab), when the interaction takes place in a near critical fluid medium with account for the omnipresent van der Waals forces (vdWF). Here we consider the case of complete wetting of the objects surfaces by the medium, due to strong adsorbing local surface potentials, exerted by thin solid coating films. The influence of the bulk inner regions of the particle and the slab on the constituents of the fluid results in long-ranged competing dispersion potentials. As a consequence from the critical fluctuations of the medium, the system experiences an additional effective interaction, traditionally termed critical Casimir force (CCF). The forces of interest are evaluated numerically from integral expressions obtained utilizing general scaling arguments and mean-field type calculations in combination with the so-called "surface integration approach" (SIA). Within the scenario considered here, this technique is applicable if one has knowledge of the forces between two parallel semi-infinite plates, confining in between some fluctuating fluid medium characterized by its temperature $T$ and chemical potential $\mu$. It is demonstrated that for a suitable set of particle-fluid, slab-fluid, and fluid-fluid coupling parameters the competition between the effects due to the coatings and the core regions of the objects result, when one changes $T$ or $\mu$, in {\it sign change} of the NF acting between the ellipsoid and the slab.
\end{abstract}
\section{INTRODUCTION}\label{sec:Intro}
When a confined fluid medium is at the vicinity of its bulk critical point, as first suggested by Fisher and de Gennes  \cite{FG78}, an additional component adds up to the already acting forces if any, resulting from the correlated critical fluctuation of the medium. This new contribution is of a long-ranged character and strongly depends on the boundary conditions which the confining objects, immersed in the medium, impose on it at their surfaces. Since this fluctuation-induced force depends additionally only on some gross features of the fluid medium \cite{K94,BDT2000} it can be treated as {\it universal} in nature. One can make a parallel with the force between neutral bodies due to the quantum and temperature fluctuations of the charge distributions in them, i.e., of the electromagnetic field, which force is known today under the general name of quantum electrodynamic (QED) Casimir force \cite{C48,CP48}. Because of this analogy it became customary to term the fluctuation part of the net force {\it critical Casimir force} \cite{K94,BDT2000}.

In a medium where both quantum and thermodynamical fluctuations are present, $F_{\rm net}$ can be thought as a sum of a regular background $F_{{\rm net}}^{\rm (reg)}$ and a singular $F_{{\rm net}}^{{\rm (sing)}}$ contributions. The first depends in an analytic way on the parameters characterizing the medium, whereas the latter emerges due to the critical fluctuations of the medium, respectively. Therefore
\begin{equation}\label{eq:decomp}
F_{{\rm net}}=F_{{\rm net}}^{\rm (reg)}+F_{{\rm net}}^{{\rm (sing)}}.
\end{equation}
It is normal to equate
\begin{equation}\label{eq:def_Casimir}
F_{{\rm Cas}}\equiv F_{{\rm net}}^{{\rm (sing)}}(L|T,\mu)\ \ \ \mbox{and}\ \ \ F_{{\rm vdW}} \equiv F_{{\rm net}}^{\rm (reg)}(L|T,\mu),
\end{equation}
since the dispersion van der Waals interactions, ubiquitous for any system, are not influenced by the thermodynamic fluctuations.

Following Refs. \cite{VD2015,VaDa2017,Va2018}, here we are going to be interested in the simultaneous manifestation of both the critical Casimir force (CCF) and the van der Waals force (vdWF), which gives rise to the {\it net} force $F_{\rm net}$ between an ellipsoidal in shape particle and a thick smooth plate, say the handling arm of a micro-gripper -- Figure 1({\bf b}). It will be demonstrated that by proper choice of the materials (cores) of the colloid particle and the plate it is indeed possible to achieve control over the net force (NF) by simply changing the temperature $T$ and chemical potential $\mu$ of the fluid medium. So far there is only one article \cite{KHD2009}, which addresses the geometry of the particle studied here. Moreover, the interactions accounted there are of pure short-range type.

The content of the article is arranged as follows. In the following subsection we recall and comment on the finite-size behaviour of the CCF, vdWF and NF when they act between pair of parallel plates. Here we also identify the main coupling parameters characterizing the dispersion interactions in the systems, and give the general expressions used to calculate the CCF and NF. Next, the "surface integration approach" (SIA), within which we study the commented spectrum of forces, is introduced. The exact equations used to evaluate numerically the investigated forces are derived in the last section, which ends with a discussion on the observed behaviour of the NF.
\subsection{Influence of the dispersion forces on the thermodynamic Casimir one in nonpolar fluid systems with film geometry}\label{subsec:Theoretical}
If an object of certain shape is introduced in a medium, it alters its thermodynamic behaviour in such a way that any quantitative occurrence shows dependance on the penetration depth of the symmetry breaking effect into the volume. The range to which this effect is felt within the system depends both on the scope of the interactions and on that of the correlations between the fluctuations in the fluid, which mediate the interactions between the bodies.
\begin{figure}[h!]
	\centering
	\includegraphics[width=\textwidth]{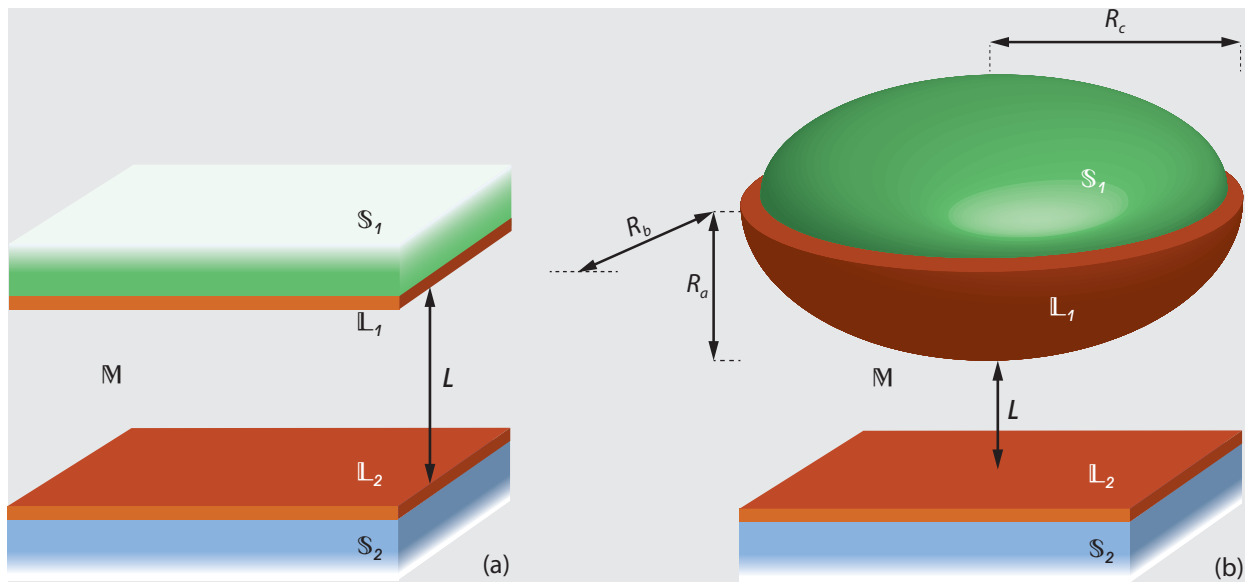}
	\caption{Schematic depiction of the considered fluid systems: $\mathbf{(a)}$ pair of parallel plates and $\mathbf{(b)}$ ellipsoidal particle with half-radii $\mathrm{\mathbf{R}}=\{R_{a},R_{b},R_{c}\}$ above a plate. The interacting objects are assumed immersed in some fluid medium $\mathbb{M}$ -- a nonpolar one-component fluid or a binary mixture composed out of the molecules of some nonpolar liquids $A$ and $B$, which is close to its critical/demixing point. The minimal separation between the interacting objects is denoted by $L$. The substances composing the objects are denoted by $\mathbb{S}_{1}$ and $\mathbb{S}_{2}$, coated by thin layers of some other substances $\mathbb{L}_{1}$ and $\mathbb{L}_{2}$, respectively. The fluid medium is considered embedded on a lattice (with cell length $a_{0}$) in which, some nodes are occupied by a particle and others are not (for a simple one-component fluid) -- thus depicting the "liquid" and "gas" states respectively at some values of the temperature $T$ and chemical potential $\mu$ of the fluid, or some of the nodes are occupied by a molecule from the substance $A$ (the "liquid" state) and the rest are occupied by the molecules belonging to the species $B$ ("gas" state). The surfaces of the interacting objects impose on the fluid medium boundary conditions of strong adsorption, the so-called $(+,+)$ boundary conditions, i.e., the nearest to the coating substances layers are entirely occupied by the particles of the one-component fluid or if the medium is a binary liquid mixture -- by the particles of one of its components. The bulk phase (core) of the objects, on the other hand, influence the fluid by long-range {\it competing} dispersion potentials.}
	\label{fig:scheme}
\end{figure}
The range of the correlations is set by the correlation length $\xi$ of the so-called \textit{order parameter} of the medium. This quantity becomes large, and theoretically diverges, at the vicinity of the bulk critical point $(T_c,\mu_c)$: $\xi(T\to T_c^{+},\mu=\mu_c)\simeq \xi_0^{+}t^{-\nu}$, where $t=(T-T_c)/T_c$, and $\xi(T=T_c,\mu\to\mu_c)\simeq \xi_{0,\mu} |\Delta\mu/(k_B T_c)|^{-\nu/\Delta}$, where $\Delta \mu=\mu-\mu_c$. Here $\nu$ and $\Delta$ are the usual critical exponents which, for classical fluids, are those of the three-dimensional Ising model, and $\xi_0^{+}$ and $\xi_{0,\mu}$ are the corresponding nonuniversal amplitudes of the correlation length along the $t$ and $\mu$ axes. When $\xi$ becomes comparable to the characteristic dimension of the system, say the separation $L$ between the objects, the size dependence of the thermodynamic functions enters into the thermodynamic potentials through the ratio $L/\xi$, and takes a scaling form given by the finite-size scaling theory \cite{BDT2000}.

Let us consider the system depicted on Figure 1({\bf a}). Upon approaching the vicinity of the bulk critical point (critical region) of $\mathbb{M}$, following Ref. \cite{VD2015} for the occurring force $f_{\rm net}^{\parallel}(L|T,\mu)$, per cross section area $\mathcal{A}$ {\it \underline {and}} $k_{B}T\equiv\beta^{-1}$, is customary to write the following expression
\begin{equation}\label{CasimirF_l}
f_{\mathrm{net}}^{\parallel}(L|T,\mu)\simeq L^{-d}X_{\rm crit}^{\parallel}\left[x_{t},x_{\mu},x_{l},\left\{x_{s_{i}},i=1,2\right\},x_{g}\right]
+(\sigma-1)\beta H_A(T,\mu) L^{-\sigma}\vartheta^{\sigma-d}.
\end{equation}
Here $d$ is the dimensionality of the system, $\sigma$ -- an exponent describing the decay of the dispersion interactions, $\vartheta$ is a microscopic length introduced for dimensional reasons (see the text below Eq. (9) in Ref. \cite{VaDa2017}), $H_{A}$ is the Hamaker term, whose dependence from the temperature and chemical potential is given by the so-called Hamaker constant \cite{P2006,I2011} (for details see the Appendix in Ref. \cite{VD2015}), $X_{\rm crit}^{\parallel}$ is dimensionless, universal scaling function, $x_{t}=t\left(L/\xi_{0}^+\right)^{1/\nu}$ and $x_{\mu}=\beta\Delta\mu\left(L/\xi_{0,\mu}\right)^{\Delta/\nu}$ are the temperature and field relevant (in renormalization group sense) scaling variables, whereas $x_{l}\propto\lambda$, $x_{s_{i}}\propto s_{i}$ and $x_{g}$ are irrelevant scaling variables (see the text below Eq. (2.6) in Ref. \cite{VD2015}) associated with the dispersion interactions in the system. The latter are introduced in the theory through the dimensionless coupling constants (see Section III in Refs. \cite{VD2015,VaDa2017})
\begin{equation}\label{s_def_l}
s_{i}=\frac{1}{2}\beta G(d,\sigma)\left(\rho_{s_{i}}J^{s_{i},l}-\rho_c J^l\right)\ \ \ \mbox{and}\ \ \ \lambda=J^{l}/J_{\mathrm{sr}}^{l},\ \ \ \mbox{with}\ \ \ \ G(d,\sigma)=4\pi^{(d-1)/2}\frac{\Gamma\left(\frac{1+\sigma}{2}\right)}{\sigma\Gamma\left(\frac{d+\sigma}{2}\right)}.
\end{equation}
In the expressions shown in Eq. (\ref{s_def_l}), $J$ and $J_{\rm sr}$ designate the long- and, respectively, the short-ranged components of the London-van der Waals potentials, between the constituents of the core regions $\mathbb{S}_{i}$ and the fluid $(s_{i},l)$ and within the medium itself $(l)$; $\rho_{s_{i}}$ and $\rho_{c}$ are the number densities of $\mathbb{S}_{i}$ and the critical one of $\mathbb{M}$, respectively, in units $a_{0}^{-d}$.

According to the scaling hypothesis of the CCF one expects that near the bulk critical point
\begin{equation}\label{scaling_function}
f_{\rm Cas}^{\parallel}(L|T,\mu)=L^{-d}X_{\mathrm{Cas}}^{\parallel}(x_t,x_\mu,\cdots),
\end{equation}
where $X_{\mathrm{Cas}}^{\parallel}$ is a scaling function, that for large enough $L$ with fixed $x_t={\cal O}(1)$ and  $x_\mu={\cal O}(1)$ approaches the scaling function of the short-ranged system $X_{\mathrm{Cas, sr}}^{\parallel}(x_t,x_\mu)$ (for details see Eqs. (2.12) and (4.10) in Ref. \cite{VD2015}). From Eqs. (\ref{eq:decomp}) and (\ref{eq:def_Casimir}), together with Eq. (\ref{CasimirF_l}) follows that $X_{\mathrm{Cas}}^{\parallel}$ is proportional to the sum of $X_{\rm crit}^{\parallel}$ and the singular part of the Hamaker term $H_{A}^{({\rm sing})}(T,\mu)$. The last implies that in order to determine the CCF in systems with dispersion interaction one has to decompose the contribution captured through the Hamaker term in a singular and a regular parts, i.e.
\begin{equation}
\label{eq:Hamaker_decomposition}
H_{A}(T,\mu)=H_{A}^{({\rm reg})}(T,\mu)+H_{A}^{({\rm sing})}(T,\mu).
\end{equation}
Thus, for $d=\sigma$ one has
\begin{equation}\label{scaling_function_Casimir}
f_{\rm Cas}^{\parallel}=L^{-d}\left[X_{\mathrm{crit}}^{\parallel}+(d-1)\beta H_A^{(\rm sing)}\right],
\end{equation}
while $f_{\rm vdW}^{\parallel}$ equals to the second term of Eq. (\ref{CasimirF_l}), with $H_{A}\equiv H_{A}^{\rm (reg)}$.

In what follows, we are going to present results for the CCF, vdWF and NF in the cases of ellipsoid-plate system, utilizing the knowledge gained from studies of the corresponding interactions between parallel plates.
\section{THE SURFACE INTEGRATION APPROACH AND ITS APPLICATION FOR THE ELLIPSOID-PLATE SYSTEM}\label{sec:SIA}
In 1934 the soviet scientist B. Derjaguin was the first to propose an approach \cite{D34} for calculating geometry dependent interactions in systems where at least one of the objects has a nonplanar geometry. In particular, this approach focuses on relating the interaction force/potential between two gently curved colloidal particles with the knowledge for that between a pair of parallel plates $f^{\parallel}_{{\cal A}}$. An important feature of this technique is that it is \textit{only} applicable if the separation distance between the interacting objects is {\it much smaller} than their geometrical characteristics. In order to overcome this inconvenient condition the co-called "surface integration approach" (SIA) reported in Ref. \cite{DV2012} was developed. Here we must note that both DA and SIA are strictly valid if the interactions involved can be described by pair potentials.

Within this new technique the interaction force between an object (say a colloid particle) $B\equiv\{(x,y,z),(x,y,z)\in B\}$ of arbitrary shape $S(x,y)=z$ and a flat surface bounded by the $(x,y)-$plane of a Cartesian coordinate system, is determined by subtracting from the contributions stemming from the surface regions $A_S^{\rm to}$ of the particle that "face towards" the plane those from the regions $A_S^{\rm away}$ that "face away" from it, i.e.,
\begin{equation}\label{SIAgeneralsimple}
F^{B,|}_{\rm SIA}(L) =\int_{A_{S}^{\rm to}}\int  f^{\parallel}_{{\cal A}}[{\rm S}(x,y)] \mathrm{d}x \mathrm{d}y-\int_{A_{S}^{\rm away}}\int  f^{\parallel}_{{\cal A}}[{\rm S}(x,y)] \mathrm{d}x \mathrm{d}y.
\end{equation}
Here $A_S^{\rm to}$ and $A_S^{\rm away}$  are the projections of the corresponding parts of the surface of the body on the $(x,y)-$plane, with $A_S=A_S^{\rm to}\bigcup A_S^{\rm away}$. The expression Eq. (\ref{SIAgeneralsimple}) also takes into account that the force on a given point of $S$ is {\it along the normal} to the surface at that point (for details see Section 2 in Ref. \cite{DV2012}). Therefore one can make use of the SIA to calculate the CCF, since under the assumption for mechanical equilibrium of the particle in the fluid, this force acts at any point of $S$ along the normal to the surface. Thus, one can get a reasonably good approximation to the effect of that force by keeping just the integration over the part of the surface of the body that faces the plane. This leads to
\begin{equation}\label{SIA_Casimir}
\beta F^{B,|}_{{\rm Cas},{\rm SIA}}(L) =\int_{A_{S}^{\rm to}}\int  f^{\parallel}_{\rm Cas}[{\rm S}(x,y)] \mathrm{d}x \mathrm{d}y.
\end{equation}

In what follows we are going to make use of the SIA to estimate the vdWF and the CCF between a planar substrate and an ellipsoidal particle with half-radii $\mathrm{\mathbf{R}}=\{R_{a},R_{b},R_{c}\}$ and center positioned at a point $(0,0,z_{c})$ with respect to the substrate's surface, where $z_{c}>\max\{R_{a},R_{b},R_{c}\}$ and $R_{c}>R_{b}>R_{a}$. The shape of the particle allows two cases to be considered:
\begin{description}
  \item[(i)]  when the distance $z_{c}=L_{c}+R_{c}$ is fixed;
  \item[(ii)] when the distance $L_{m}$ of closest approach between the surfaces of the interacting objects is fixed.
\end{description}
Here $L_{c}$ is the separation between the particle and the substrate when $R_{c}$ points toward the $(x,y)-$plane.

Hence, using Eq. (\ref{SIAgeneralsimple}) one can write the following general expression in case \textbf{(i)}
\begin{equation}\label{eq:fixed_c_ElP}
F_{\rm SIA}^{\mathrm{El},|}(L_{c})=2\pi R_{c} D \int_{L_{\min}}^{L_{\max}}\left(1-\frac{z-L_{c}}{R_{c}}\right)f_{{\cal A}}^{\parallel}(z)\mathrm{d}z,
\end{equation}
whereas in case \textbf{(ii)}
\begin{equation}\label{eq:fixed_m_ElP}
F_{\rm SIA}^{\mathrm{El},|}(L_{m})=2\pi r_{\rm eff} D \int_{L_{m}}^{L_{m}+2r_{\rm eff}}\left(1-\frac{z-L_{m}}{r_{\rm eff}}\right)f_{{\cal A}}^{\parallel}(z)\mathrm{d}z.
\end{equation}
In Eqs. (\ref{eq:fixed_c_ElP}) and (\ref{eq:fixed_m_ElP}) the following notations are introduced
\begin{equation}\label{eq:notations}
r_{\rm eff}=\sqrt{R_{a}^{2}\sin^{2}\theta+\cos^{2}\theta\left(R_{c}^{2}\cos^{2}\phi+R_{b}^{2}\sin^{2}\phi\right)},\ D=\frac{R_{a}R_{b}R_{c}}{r_{\rm eff}^{3}},\ L_{\max,\min}=z_{c}\pm r_{\rm eff}.
\end{equation}
In the definition for $r_{\rm eff}$, $\phi$ and $\theta$ denote two of the Euler angles to which, in general, the particle can be orientated with respect to a right-handed Cartesian coordinate system, attached to the surface of the planar substrate (see Figure 4 in Ref. \cite{DV2012}). Substituting $f_{\rm vdW}^{\parallel}$ in the above expressions, for the regular (van der Waals) contribution to the NF between an ellipsoid and a plate one obtains in cases \textbf{(i)} and \textbf{(ii)} the following
\begin{equation}\label{eq:Ell_c}
F_{\rm vdW,SIA}^{\mathrm{El},|}(L_{c})=\frac{2\pi H_{A}^{\rm (reg)}\vartheta^{\sigma-3}D}{\sigma-2}\left[\frac{z_{c}(\sigma-2)-L_{\min}(\sigma-1)}{L_{\min}^{\sigma-1}}-
\frac{z_{c}(\sigma-2)-L_{\max}(\sigma-1)}{L_{\max}^{\sigma-1}}\right],
\end{equation}
and
\begin{equation}\label{eq:Ell_m}
F_{\rm vdW,SIA}^{\mathrm{El},|}(L_{m})=\frac{2\pi H_{A}^{\rm (reg)}\vartheta^{\sigma-3}D}{\sigma-2}\left[\frac{r_{\rm eff}(\sigma-2)-L_{m}}{L_{m}^{\sigma-1}}+
\frac{L_{m}+r_{\rm eff}\sigma}{(L_{m}+2r_{\rm eff})^{\sigma-1}}\right],
\end{equation}
respectively.

The corresponding expressions in $d=3$ for the CCF arising between the studied objects, within the two considered cases, follow directly from the substitution of Eq. (\ref{scaling_function}) in Eq. (\ref{SIA_Casimir}), and have the form
\begin{equation}\label{eq:CasEl_c}
\beta F_{\rm Cas, SIA}^{{\rm El},|}(L_{c})= 2\pi R_{c} D \int_{L_{\min}}^{(L_{\max}-L_{\min})/2}\left(1-\frac{z-L_{c}}{R_{c}}\right)\frac{1}{z^{3}}X_{\mathrm{Cas}}^{\parallel}[x_t(z),x_\mu(z),\cdots]{\rm d}z\equiv 2\pi R_{c}D X_{\mathrm{Cas,SIA}}^{{\rm El},|}(L_{c}|x_t,x_\mu,\cdots),
\end{equation}
and
\begin{equation}\label{eq:CasEl_m}
\beta F_{\rm Cas, SIA}^{{\rm El},|}(L_{m})= 2\pi r_{\rm eff} D \int_{L_{m}}^{L_{m}+r_{\rm eff}}\left(1-\frac{z-L_{m}}{r_{\rm eff}}\right)\frac{1}{z^{3}}X_{\mathrm{Cas}}^{\parallel}[x_t(z),x_\mu(z),\cdots]{\rm d}z\equiv 2\pi r_{\rm eff} D X_{\mathrm{Cas,SIA}}^{{\rm El},|}(L_{m}|x_t,x_\mu,\cdots).
\end{equation}
These two equations are strictly valid {\it only} when: {\textbf{a)}} $\phi=\theta=0$, i.e., $R_{c}$ points toward the $(x,y)-$plane; {\textbf{b)}} $\phi=\pi/2, \theta=0$, i.e., $R_{b}$ points toward the $(x,y)-$plane and {\textbf{c)}} any value of $\phi$ if $\theta=\pi/2$, i.e., when $R_{a}$ points toward the $(x,y)-$plane. This is because the CCF originates from the {\it confinement} of the correlated fluctuations of the fluid at its critical point. When the particle is arbitrarily orientated the projected infinitesimal area lies either between two nonconcentric ellipses, or when the ellipses intersect $\mathrm{d}S$ is their disjoint area. The latter has a part which faces the plate, and hence contributes to the CCF, and such that is orientated away from it (the direction of the $z-$component ${\bf{\rm{e}}}_{z}$ of the unit normal vector $\mathbf{\mathrm{n}}_{\mathbf{\mathrm{r}}}$ coincides with that of the $z-$coordinate -- see Figure 4 in Ref. \cite{DV2012} and the text therein). Hence Eqs. (\ref{eq:CasEl_c}) and (\ref{eq:CasEl_m}) {\it can not} be used outside cases {\textbf{a)}} - {\textbf{c)}}, within which the spectrum of occurred fluctuations is restricted {\it only} between the lower half of the ellipsoid's surface and that of the plate.
\section{RESULTS}\label{sec:ResultsAndExpFeas}
Within the mean-field theory the $T$ and $\mu$ dependance of the CCF and hence the NF between any pair of objects immersed in near critical fluid medium $(T\approx T_c,\Delta\mu\approx0)$ is given by the temperature and field scaling variables $x_{t}$ and $x_{\mu}$, respectively [see the text beneath Eq. (\ref{CasimirF_l})], with $\nu=1/2$ and $\Delta=3/2$, i.e., for $d=4$. In our numerical treatment we take these variables to range in the intervals: $x_{t}\in\left[-24^{2};24^{2}\right]$ and $x_{\mu}\in\left[-24^{3};24^{3}\right]$. To study the scaling function of the CCF, the separation between the set of parallel plates is varied from 20 to 120 with step 2, from 23 to 45 with step 1, from 100 to 120 with step 1, from 120 to 220 with step 5 and from 180 to 220 with step 2. In order to demonstrate the effect of the sign change of the considered forces and in view of potential experimental realization of the predicted effects we choose the following values of the coupling parameters: $s_{1}=1.0$, $s_{2}=-0.01$ and $\lambda=2.0$. The physical argumentation of the so chosen values is explained in Section IV C of Ref. \cite{VaDa2017}.
\subsection{The van der Waals and critical Casimir forces between an ellipsoid and a plate in $d=3$ within the SIA}\label{subsec:SpPlSpSpSIAd3}
In $d=3$, considering genuine $(\sigma=d)$ van der Waals interactions between all constituents comprising the system ellipsoid-fluid-plate, Eqs. (\ref{eq:Ell_c}) and (\ref{eq:Ell_m}) become
\begin{equation}\label{eq:EllLcmsigma3}
L_{c}\beta F_{\rm vdW,SIA}^{{\rm El},|}=\left[2\beta H_{A}^{\rm (reg)}\right]\frac{4\pi\Xi_{a}\Xi_{b}\Xi_{c}}{\left(\bar{z}_{c}^{2}-\vartheta_{\rm eff}^{2}\right)^{2}}\ \ \ \ \  \mbox{and}\ \ \ \ \ L_{m}\beta F_{\rm vdW,SIA}^{{\rm El},|}=\left[2\beta H_{A}^{\rm (reg)}\right]\frac{4\pi\Xi_{a}'\Xi_{b}'\Xi_{c}'}{\left(1+2\vartheta_{\rm eff}'\right)^{2}},
\end{equation}
respectively. Here the following notations are introduced: $\Xi\equiv R/L$, $\vartheta_{\rm eff}\equiv r_{\rm eff}/L$, with $L$ interpreted either as $L_{c}$ or $L_{m}$ depending on the considered case, $\bar{z}_{c}\equiv 1+\Xi_{c}$. The quantities normalized by $L_{m}$ contain an additional prime symbol $(')$. Accordingly for the CCF from Eqs. (\ref{eq:CasEl_c}) and (\ref{eq:CasEl_m}) one can write
\begin{equation}\label{eq:CasEl_cm_dim}
L_{c}\beta F_{\rm Cas, SIA}^{{\rm El},|}=2\pi \frac{\Xi_{a}\Xi_{b}\Xi_{c}^{2}}{\vartheta_{\rm eff}^{3}\left(\bar{z}_{c}-\vartheta_{\rm eff}\right)^{2}} X_{\mathrm{Cas,SIA}}^{{\rm El},|}(L_{c})\ \ \ \ \  \mbox{and}\ \ \ \ \ L_{m}\beta F_{\rm Cas, SIA}^{{\rm El},|}=2\pi \frac{\Xi_{a}'\Xi_{b}'\Xi_{c}'}{\vartheta_{\rm eff}^{'2}} X_{\mathrm{Cas,SIA}}^{{\rm El},|}(L_{m}),
\end{equation}
where
\begin{equation}\label{eq:XCasElPl}
X_{\mathrm{Cas,SIA}}^{{\rm El},|}=\int_{1}^{\Lambda_{1}}I(\bar{z}|\Lambda_{1})X_{\rm Cas}^{\parallel}[\mathfrak{a}_{1}(\bar{z})]{\rm d}\bar{z}+\sum_{i=j+1}^{j_{\max}-1}\int_{\Lambda_{i-1}}^{\Lambda_{i}}I(\bar{z}|\Lambda_{i})X_{\rm Cas}^{\parallel}[\mathfrak{a}_{i}(\bar{z})]{\rm d}\bar{z}
+\int_{\Lambda_{\max}}^{1+U_{\rm lim}}I(\bar{z}|\Lambda_{\max})X_{\rm Cas}^{\parallel}[\mathfrak{a}_{\max}(\bar{z})]{\rm d}\bar{z},
\end{equation}
with $\Lambda_{j}\equiv L_{j}/L,\ j=1,j_{\max}$; $\bar{z}\equiv z/L$ is dimensionless variable; $L_{\max}\equiv L_{j_{\max}}$ is the largest system for which numerical data are available, the arguments of the scaling function $[\mathfrak{a}_{j}(\bar{z})]\equiv[x_{t}(L_{z;j}),x_{\mu}(L_{z;j}),...]$, $L_{z;j}\equiv \bar{z}L_{j}$ and $\theta$ is the Heaviside step function with the convention $\theta(0)=0$. For the present study $L_{\max}=220$. In Eq. (\ref{eq:XCasElPl}) with respect to case \textbf{(i)} we have
\begin{equation}\label{case_i}
L\equiv L_{\min},\ I(\bar{z}|\Lambda_{j})=\frac{1}{\bar{z}^{3}}\left[1-\frac{\bar{z}\left(\bar{z}_{c}-\vartheta_{\rm eff}\right)-1}{\Xi_{c}}\right]\theta(U_{\rm lim}-\Lambda_{j}),\ U_{\rm lim}=\frac{\vartheta_{\rm eff}}{\bar{z}_{c}-\vartheta_{\rm eff}},
\end{equation}
while when one considers case \textbf{(ii)}
\begin{equation}\label{case_ii}
L\equiv L_{m},\ I(\bar{z}|\Lambda_{j})=\frac{1}{\bar{z}^{3}}\left(1-\frac{\bar{z}-1}{\vartheta_{\rm eff}'}\right)\theta(U_{\rm lim}-\Lambda_{j}),\ U_{\rm lim}=\vartheta_{\rm eff}'.
\end{equation}

In Eq. (\ref{eq:XCasElPl}) the critical component $X_{\rm crit}^{\parallel}$ of the scaling function $X_{\rm Cas}^{\parallel}$ calculated within the mean-field theory is normalized by the procedure described in-and-out in Section V of Ref. \cite{VD2015}, so that it contributes properly to the CCF and hence to the net interaction force when $d<4$. The need of such normalization is explained in details in Ref. \cite{DSD2007} (see there Sections IV.A.1 and IV.A.3).

After presenting the mathematical means to calculate the CCF and vdWF, we now pass to the detailed discussion of the results and argumentation of the experimental feasibility of the parameters used in the model calculations utilizing the SIA.
\subsection{Discussion and concluding remarks}\label{subsec:Discussion}
\begin{figure*}[h!]
  \centering
\mbox{\subfigure{\includegraphics[height=5.1 cm]{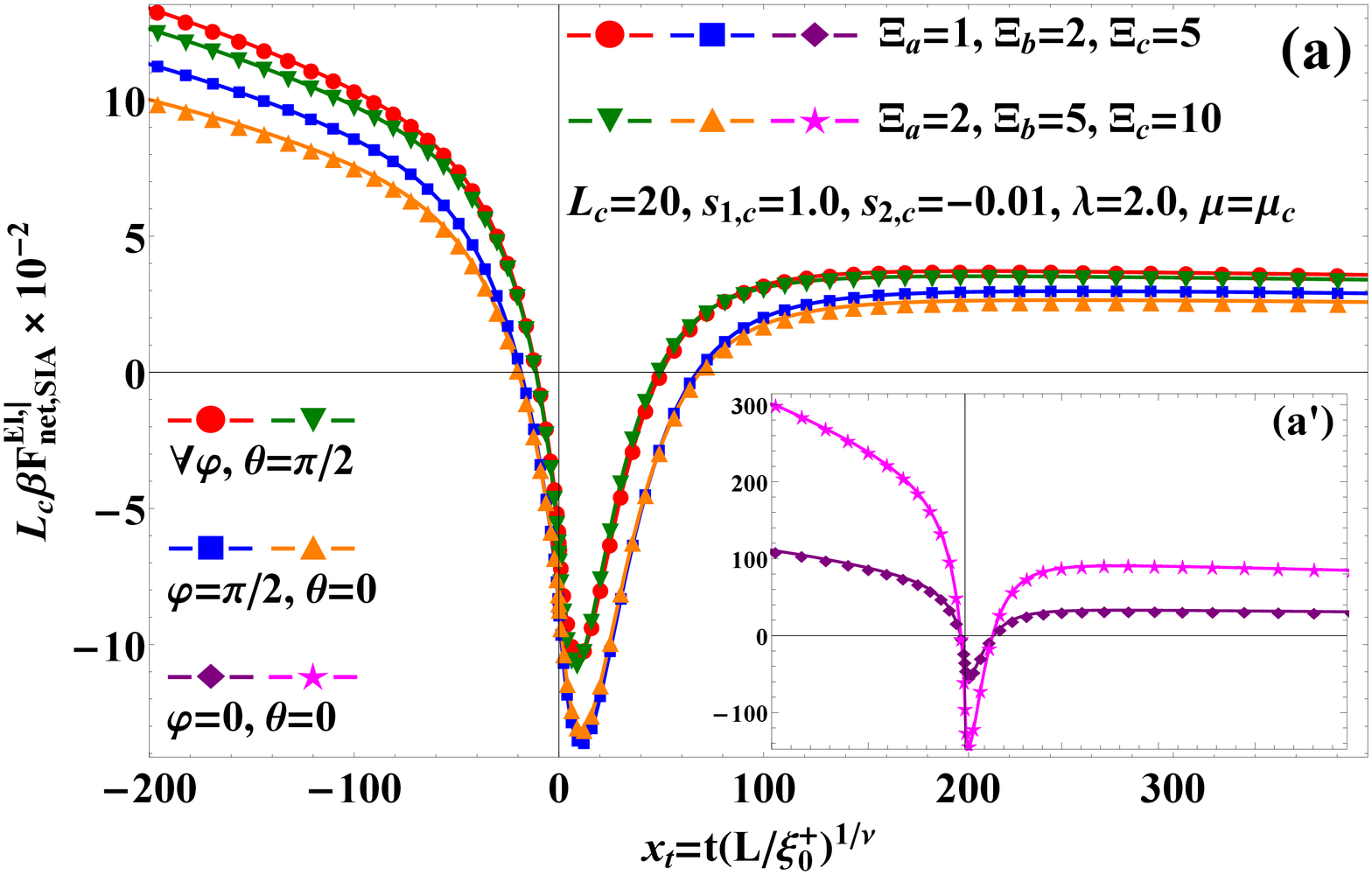}}\quad
      \subfigure{\includegraphics[height=5.1 cm]{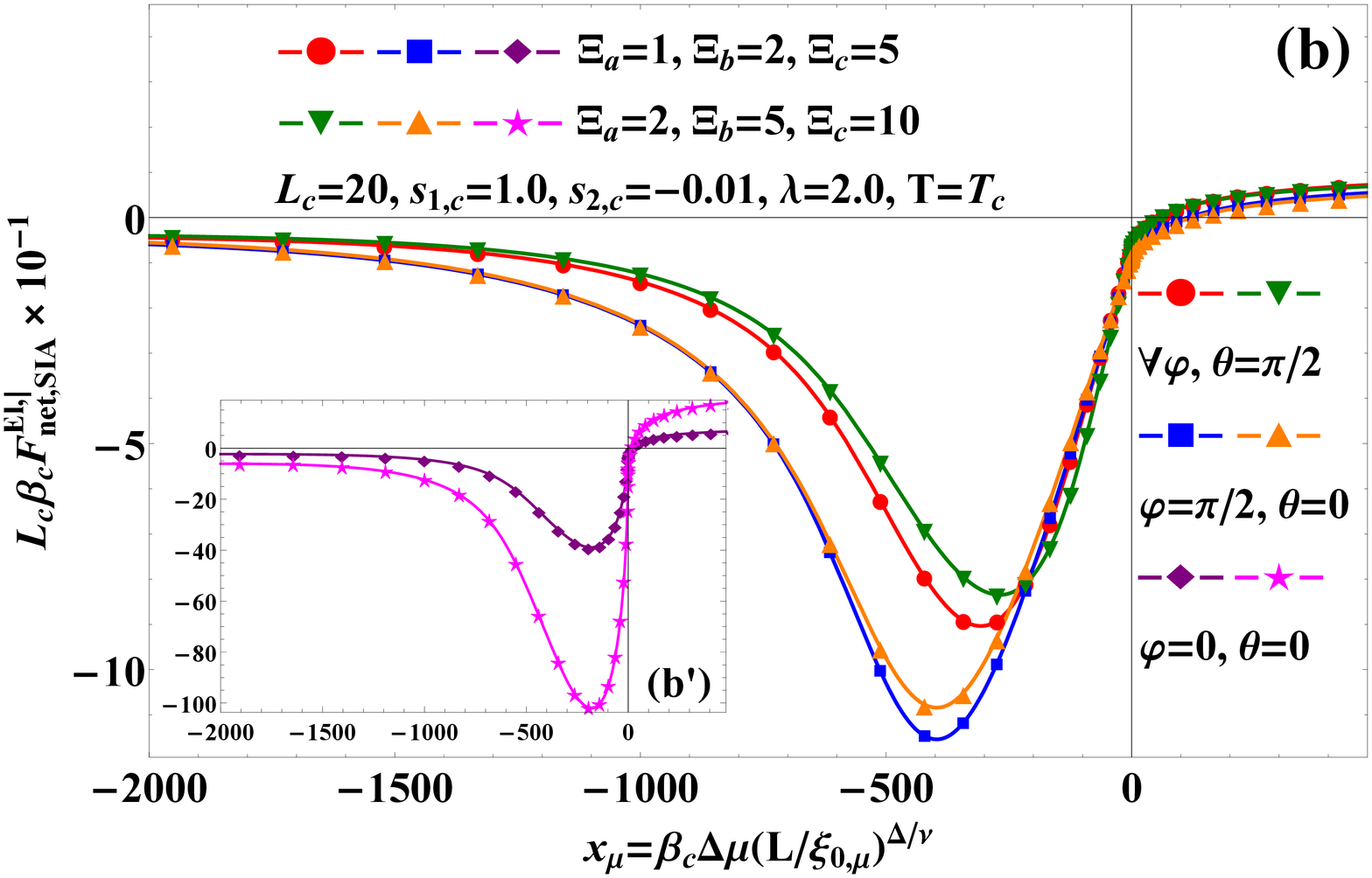}}}\\
\mbox{\subfigure{\includegraphics[height=5.1 cm]{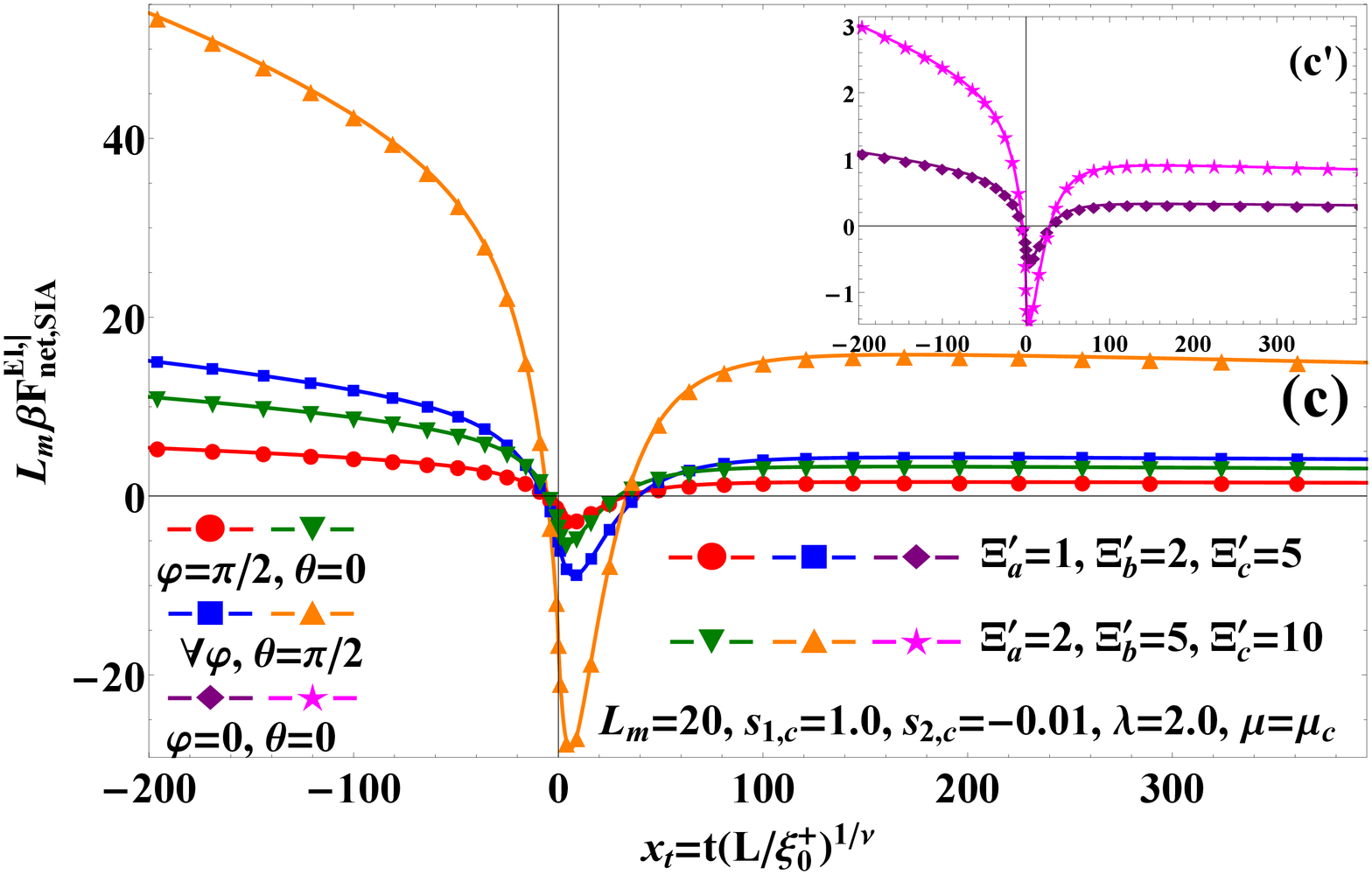}}\quad
      \subfigure{\includegraphics[height=5.1 cm]{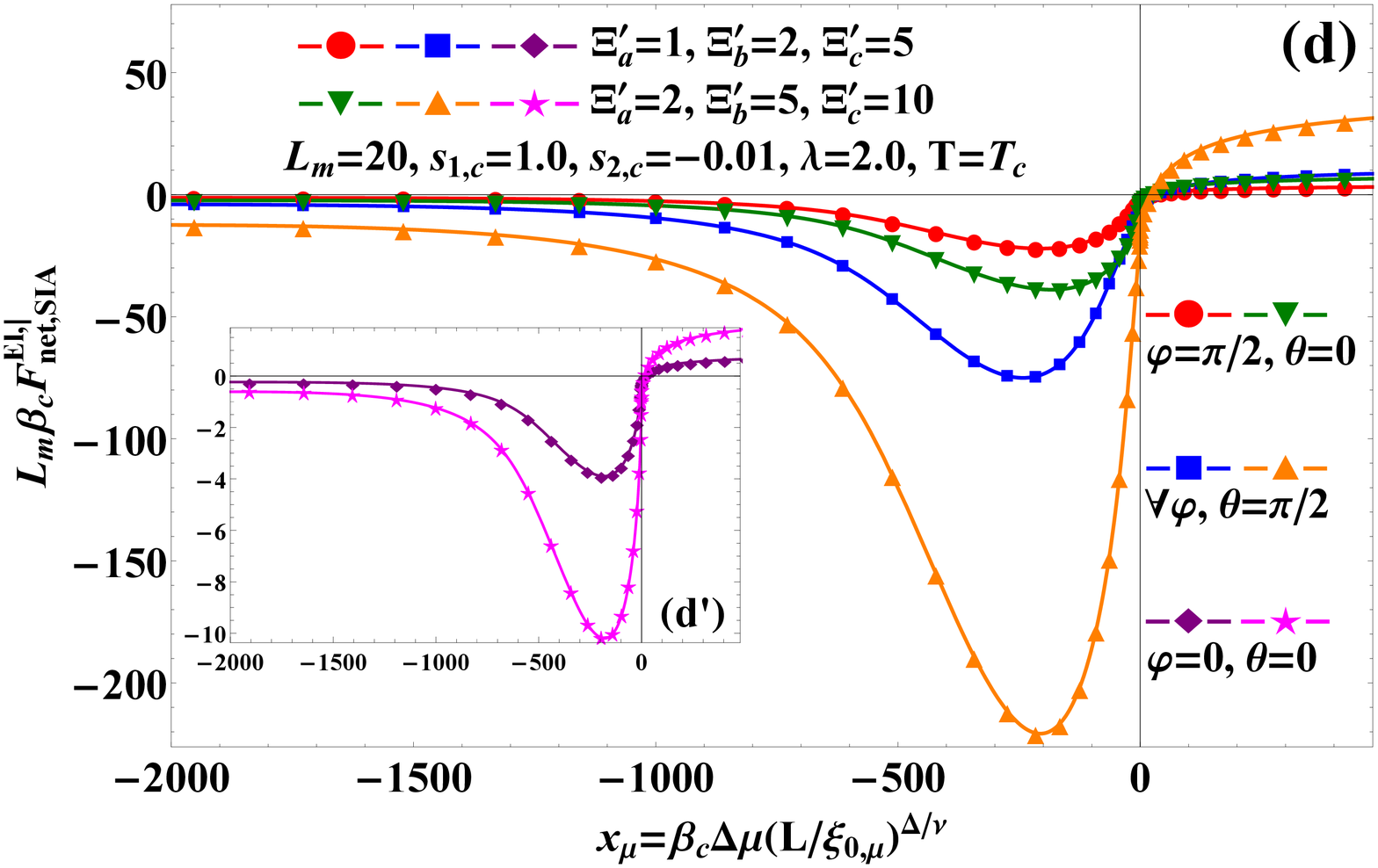}}}
  \caption{Behavior of the net force $F_{\rm net,SIA}$ (normalized by the separation $L$ and $k_{B}T$) in $d=3$ when one considers $z_{c}$ - [$\mathbf{(a)}$ and $\mathbf{(b)}$] or the minimal separation $L_{m}$ - [$\mathbf{(c)}$ and $\mathbf{(d)}$] fixed. The temperature dependance at $\Delta\mu=0.0,\ (x_{\mu}=0.0)$ is depicted on $\mathbf{(a)}$ and $\mathbf{(c)}$, while that of the field one at $T=T_{c},\ (x_{t}=0.0)$ on $\mathbf{(b)}$ and $\mathbf{(d)}$. Each of the considered separations $L_{c}$ and $L_{m}$ is taken $20a_{0}$ and the parameters characterizing the interactions in the systems have values $\lambda=2.0$, $s_{2,c}=-0.01$ and $s_{1,c}=1.0$.}
  \label{fig:total_forces}
\end{figure*}
In the present section for physical reasons related to future experimental studies we will assume that the ellipsoidal particle has either ruthenium (Ru) or platinum (Pt) core, while the planar substrate is made either of carbon (C) or silica (SiO$_{2}$) aerogels. The contact surface of any of the interacting objects is considered coated by monolayer of lead (Pb) or thallium nitride (TlN) to ensure the (+,+) boundary conditions. Finally, we complete the so assembled system by adopting xenon (Xe) as the critical fluctuating medium in which the interacting objects are immersed. For detail comment and argumentation on the choice of the presented substances the reader can refer to Ref. \cite{VaDa2017}.

Some results about the behavior of the net force in an ellipsoid-plate system are presented on Figure \ref{fig:total_forces}. The effect of the particle size is included in the study through the consideration of particles with dimensions: $P_{1}:\{\Xi_{a},\Xi_{b},\Xi_{c}\}=\{1,2,5\}$ and $P_{2}:\{\Xi_{a},\Xi_{b},\Xi_{c}\}=\{2,5,10\}$. For the sake of simplicity we choose $L_{c}=L_{m}=20a_{0}\simeq12\ {\rm nm}$. The analysis of the numerical data showed that the critical component of the Casimir force is \textit{negative} for any value of $x_{t}$ and $x_{\mu}$, i.e., it corresponds to \textit{attraction} between the ellipsoid and the plate. On the other hand at $\Delta\mu=0$, $H_{A}^{\rm (sing)}$ has \textit{positive} values for $T<T_{c}$, i.e. in the liquid phase of $\mathbb{M}$, and is \textit{zero} for $T\geq T_{c}$, since it is proportional to the bulk order parameter of the medium. Therefore, the behavior of Casimir interaction $F_{\rm Cas,SIA}^{{\rm El},|}=F_{\rm sing,SIA}^{{\rm El},|}+F_{\rm crit,SIA}^{{\rm El},|}$ coincides with that of its critical component above $T_{c}$, but one is able observes a single \textit{sign change} in the "liquid" phase of the fluid. The superposition between the $F_{\rm vdW,SIA}^{{\rm El},|}$ and $F_{\rm sing,SIA}^{{\rm El},|}$, leads to interaction which is positive for any temperature and shows explicit decrease with the increase of the temperature like $T^{-1}$ (see additionally Eqs. (33) and (34) in Ref. \cite{Va2018}). As a result $F_{\rm net,SIA}^{{\rm El},|}$ becomes \textit{repulsive} outside the critical region, \textit{changes sign twice}, after which turns \textit{attractive} only near the critical point. The nature of this observation is dictated by the increased correlations between the fluctuations in the fluid close to $T_{c}$. The so described behaviour is well illustrated on subfigured (\textbf{a}) and (\textbf{c}).

Now, at $T=T_{c}$ as a function of $x_{\mu}$, $F_{\rm sing,SIA}^{{\rm El},|}$ is \textit{attractive} in the "gas" phase of the fluid, i.e., for $x_{\mu}<0$ and \textit{repulsive} otherwise with an infinite slope at the critical point, where a sign change occurs. Hence, the CCF \textit{changes sign} once in the "liquid" phase ($x_{\mu}>0$) and its minimum increases in magnitude slightly in comparison to $F_{\rm crit,SIA}^{{\rm El},|}$. Since $H_{A}^{\rm (reg)}=\mathrm{const.}>0$  does not depend on the chemical potential, when the corresponding vdWF is added to $F_{\rm sing,SIA}^{{\rm El},|}$, the resultant interaction "shift up" and change sign for $\mu<\mu_{c}$. Therefore, as presented on subfigured (\textbf{b}) and (\textbf{d}), the net force exhibits behavior similar to that of $F_{\rm Cas,SIA}^{{\rm El},|}$, but now the sign change occurs closer to the critical point.

With respect to the mutual orientation between the interacting objects, when $z_{c}$ is fixed, as expected $F_{\rm net,SIA}^{{\rm El},|}$ is maximal [subfigured (\textbf{a}') and (\textbf{b}')] at any $T$ and $\mu$ when $\varphi=\theta=0$, and minimal for any value of $\varphi$ and $\theta=\pi/2$. Additionally, one observes that in the critical region the net interactions for $P_{1}$ and $P_{2}$ are very close both quantitatively and qualitatively, when the half-radii $R_{a}$ or $R_{b}$ are orientated towards the plate. This observation can be attributed to the increased minimal separation $L_{\min}$ [entering explicitly in Eq. (\ref{case_i})] in these two cases in comparison to the one when $R_{c}$ is towards the plate. Hence irrespective of the initial orientation, if one is able to maintain $z_{c}$ fixed, the fluctuations in the system are going to drive the particle in a position where $R_{c}\perp(x,y)-$plane.

In contrast to the above, for fixed $L_{m}$, the behavior of $F_{\rm net,SIA}^{{\rm El},|}$ with respect to its angular dependance is in reverse -- the force is maximal for any value of $\varphi$ and $\theta=\pi/2$ and minimal when $\varphi=\theta=0$ [subfigured (\textbf{c}') and (\textbf{d}')]. One also observes that in each of the three orientations the minimum of the NF is deeper for $P_{2}$ than for $P_{1}$. This is easily understood having that: $\mathbf{1)}$ the confining surface of $P_{2}$ is bigger than that of $P_{1}$ and $\mathbf{2)}$ when $R_{a}\perp(x,y)-$plane maximal part of it is close to the plates surface, thus minimazing the separation effect. Therefore, this is the equilibrium orientation which the system will strive to achieve.

Although the magnitude of the NF may seems rather negligible, a comparison with say the weight of a single particle proofs otherwise. For instance, the weight of a platinum ellipsoid in shape particle with dimensions $\{R_{a},R_{b},R_{c}\}=\{2,5,10\}\times L_{m}a_{0}$ is approximately $F_{\rm w}^{\rm El}\approx0.14\ {\rm fN}$. In the liquid phase of xenon at $T=206.4\ {\rm K}$ and $\Delta\mu=0$ one finds that $F_{\rm net,SIA}^{\rm El,|}(0,0)\approx1\ {\rm pN}$, $F_{\rm net,SIA}^{\rm El,|}(\pi/2,0)\approx4\ {\rm pN}$ and $F_{\rm net,SIA}^{\rm El,|}(\forall\varphi,\pi/2)\approx20\ {\rm pN}$. Therefore the repulsive part of the net force in one such concrete system is indeed capable of levitating a single particle or a system of such. Thus, one can indeed make use of the interplay between the quantum and thermodynamical fluctuations for governing the behavior of objects, say colloidal particles, at small distances. It can also provide a strategy for solving problems with handling, feeding, trapping and fixing of micro parts in nanotechnology resolving the issues related to sticking of the particles on the surface of the mechanical manipulator utilizing, e.g., the {\it reversible dependence} on the forces under minute changes of the temperature of the critical medium. One can perform grabbing of particles for small values of $x_t$, where the force is attractive and release them at a given spacial position after slightly increasing or decreasing of temperature achieving in that way a value of $x_t$ with a repulsive NF.

\section*{Acknowledgements}
The authors gratefully acknowledges the financial support via Contract No. DN 02/8 of Bulgarian NSF.

\nocite{*}
\bibliographystyle{aipnum-cp}%

\end{document}